\documentclass{article}
\usepackage{amsmath}
\usepackage{amscd}
\usepackage{graphicx}
\usepackage{latexsym}
\usepackage{amsmath,amsthm,amsfonts,amssymb,cite%
}
\usepackage{latexsym}
\usepackage{amsmath}

\theoremstyle{remark}

\newtheorem*{rem}{Remark}
\renewcommand\dagger*
\renewcommand\Theta{\boldsymbol\theta}
\numberwithin{equation}{section}
\begin{document}
\hoffset = -2.4truecm \voffset = -2truecm
\renewcommand{\baselinestretch}{1.2}
\newcommand{\mb}{\makebox[10cm]{}\\ }
\title{A semi-discrete Kadomtsev-Petviashvili equation and its coupled integrable system}
\author{Chun-Xia Li$^{1,2}$, St\'ephane Lafortune$^{2}$ and Shou-Feng Shen$^{3}$ \\
\\
$^{1}$ School of Mathematical Sciences,
Capital Normal University,
Beijing 100048, CHINA \\
$^2$ Department of Mathematics, College of Charleston, Charleston, SC 29401, USA \\
$^3$Department of Applied Mathematics, Zhejiang University of Technology, Hangzhou 310023, CHINA}
\date{}
\maketitle

\begin{abstract}
We establish connections between two cascades of integrable systems generated from the continuum limits of the Hirota-Miwa equation and its remarkable nonlinear counterpart under the Miwa transformation respectively. Among these equations, we are mainly concerned with the semi-discrete bilinear Kadomtsev-Petviashvili (KP) equation which is seldomly studied in literature. We present both of its Casorati and Grammian determinant solutions. Through the Pfaffianization procedure proposed by Hirota and Ohta, we are able to derive the coupled integrable system for the semi-discrete KP equation.
\end{abstract}

\vskip 0.5cm

%Mathematics Subject Classification (2000). 37K10, 35Q51.
%
%\vskip 0.5cm
%
%PACS: 02.30Ik, 05.45Yv


\section{Introduction}
In the early 1990s, Hirota and Ohta \cite{HO, HR1} developed a procedure
for generalizing equations from the KP hierarchy to produce
their coupled systems, which we now call Pfaffianization.
These Pfaffianized equations appear as coupled systems of
`un-Pfaffianized' equations and have solutions expressed in terms of Pfaffians. 
So far, this procedure has been successfully applied to the Davey-Stewartson equations, the discrete KP equation, the self-dual Yang-Mills equation, the differential-difference KP equation, the two-dimensional Toda lattice equation and the semi-discrete Toda equation, etc. \cite{GN,OHT,ONG,ZLH,HZT,LH}. Besides, the Pfaffianized KP
hierarchies have been investigated in \cite{CG}. The key points involved in this procedure are first to find solutions to an `un-Pfaffianized' bilinear equation expressed in terms of Wronskian, Casorati or
Grammian type determinants, then to construct a Pfaffian with
elements satisfying the Pfaffianized dispersion
relations and finally to seek coupled bilinear
equations satisfied by these Pfaffians by using Pfaffian identities.

As is well known, a cascade of integrable systems can be produced from the Hirota-Miwa equation under the Miwa transformation \cite{HR,MT}. In our previous work \cite{LNT}, we derived a remarkable nonlinear counterpart of the Hirota-Miwa equation which we call the nonlinear Hirota-Miwa equation. In the same way as the Hirota-Miwa equation, we were able to derive a cascade of integrable systems from the nonlinear Hirota-Miwa equation. Among the cascade of bilinear integrable systems, we are mainly concerned with a differential-difference equation with two discrete and one continuous variables. This equation is seldomly studied in literature. We call this equation the semi-discrete KP equation and denote as $\Delta^2\partial$KP for short. 
 
In this paper, we will first establish the connections between the two cascades of bilinear and nonlinear integrable systems. Then we aim to apply the Pfaffianization procedure to the $\Delta^2\partial$KP equation. In order to do so, we will first find both Casorati and Grammian determinant solutions to the $\Delta^2\partial$KP equation by employing Wronskian (Casorati) technique \cite{JS,FN1,FN2}. Then by applying the Pfaffianization procedure, we manage to derive the coupled integrable system of the $\Delta^2\partial$KP equation.

This paper is organized as follows. In Section 2, we first review some results given in \cite{LNT} and then clarify that how the continuum limits of the Hirota-Miwa equation and their nonlinear counterparts are related. In Section 3, we present both Casorati and Grammian determinant solutions to the $\Delta^2\partial$KP equation. In Section 4, the coupled integrable system for the $\Delta^2\partial$KP equation is derived by Pfaffianization. Conclusions are given in Section 5.

\section{On the continuum limits of the Hirota-Miwa equation and their nonlinear counterparts}
The well-known Hirota-Miwa equation reads as \cite{HR,MT}
\begin{eqnarray}\label{bHM}
&&a_1(a_2-a_3)T_1(\tau)T_{23}(\tau)+a_2(a_3-a_1)T_2(\tau)T_{31}(\tau)\nonumber\\
&&\qquad\qquad+a_3(a_1-a_2)T_3(\tau)T_{12}(\tau)=0,
\end{eqnarray} 
where $\tau$ is a function of discrete variables $n_1, n_2$ and $ n_3$, $T_i$ is the shift operator defined by $T_i(\tau)=\tau(n_i+1)$, $T_{ij}(\tau)=T_i(T_j(\tau))$ and $a_i$ are lattice parameters.
Let us denote $\tau(n_1,n_2,n_3)=\tau(n_1,n_2,n_3;x_1,x_2,x_3,\cdots)$ in order to introduce continuous variables $x_1,\,x_2$ and $x_3$ etc.. By resorting to the Miwa transformation
$$x_k=\sum_{i=1}^\infty\frac{a_i^k}{k}n_i,$$
one can obtain a cascade of bilinear equations by taking continuum limits of \eqref{bHM} in sequence. In the following, we denote $x_1=x, x_2=y, x_3=t$. As $a_3\rightarrow 0$, we have the semi-discrete KP equation ($\Delta^2\partial$KP) 
\begin{equation}\label{bDPKP}
(a_1-a_2)(T_1(\tau)T_2(\tau)-T_{12}(\tau)\tau)+a_1a_2D_xT_1(\tau)\cdot T_2(\tau)=0.
\end{equation}
As $a_2\rightarrow 0$, we have the differential-difference KP equation ($\Delta\partial^2$KP for short) 
\begin{equation}\label{bPDKP}
(a_1D_y+a_1D_x^2-2D_x)T_1(\tau)\cdot \tau=0.
\end{equation}
Finally we have the KP equation 
\begin{equation}\label{bKP}
(D_x^4+3D_y^2-4D_xD_t)\tau\cdot \tau=0.
\end{equation}
Here, we would like to remark that the Hirota's bilinear operators $D_x$, $D_y$ and $D_n$ \cite{HR2,HS,HRN} are defined by
\begin{eqnarray*}
   &&D_x^pD_t^q\; a\cdot b\equiv\left.\left({\partial\over\partial x}-
    {\partial\over\partial x'}\right)^p\left({\partial\over\partial t}-
    {\partial\over\partial t'}\right)^qa(x,t)b(x',t')\right|_{x'=x,t'=t},\\
  &&\exp(\delta D_n)\;a\cdot b\equiv a(n+\delta)b(n-\delta),
\end{eqnarray*}
respectively.

In \cite{LNT}, Li, Nimmo and Tamizhmani proposed a remarkable nonlinear Hirota-Miwa equation 
 \begin{eqnarray}\label{fdkp}
 &&a_1a_2(\Delta_1\Delta_2 G)(a_1\Delta_1 G-a_2\Delta_2 G)\nonumber\\
 &&\qquad+a_2a_3(\Delta_2\Delta_3 G)(a_2\Delta_2 G-a_3\Delta_3 G)\nonumber\\
 &&\qquad+a_3a_1(\Delta_3\Delta_1 G)(a_3\Delta_3 G-a_1\Delta_1 G)\nonumber\\
&&\qquad=(a_1-a_2)\Delta_1\Delta_2 G+(a_2-a_3)\Delta_2\Delta_3 G+(a_3-a_1)\Delta_3\Delta_1 G,
 \end{eqnarray} 
where $\Delta_i=a_i^{-1}(T_i-1)$ denotes the forward difference operator. Since we are considering the commutative case, the additional terms involving the commutators given in \cite{LNT} are not present here. 

In a similar way as for \eqref{bHM}, by taking continuum limits of \eqref{fdkp} under Miwa transformations, we obtained a series of nonlinear equations. They are the nonlinear semi-discrete KP equation
\begin{eqnarray}\label{nDPKP}
&&a_1a_2(\Delta_1\Delta_2G)(a_1\Delta_1G-a_2\Delta_2G)\nonumber\\
&&\quad=(a_1-a_2)\Delta_1\Delta_2G+a_2\Delta_2G_x-a_1\Delta_1G_x,
\end{eqnarray}
the nonlinear differential-difference KP equation
\begin{equation}\label{nPDKP}
a_1^2(\Delta_1G_x)(\Delta_1 G)+\Delta_1G_x-G_{xx}=\frac{1}{2}a_1\Delta_1(G_{xx}+G_y)
\end{equation}
and the nonlinear KP equation
\begin{equation}\label{nKP}
3G_{yy}+(-4G_t-6G_x^2+G_{xxx})_x=0.
\end{equation}

Actually, the two sets of equations described above are closely related through the system \cite{JN,GN2}
\begin{eqnarray}
&&U_{ij}+U_{jk}+U_{ki}=0,\label{ndKP1}\\
&&T_k(U_{ij})+T_i(U_{jk})+T_j(U_{ki})=0,\label{ndKP2}\\
&&T_k(U_{ij})U_{ki}=T_j(U_{ki})U_{ij},\label{ndKP3}
\end{eqnarray}
 where $i,\,j,\,k\in \{1,2,3\}$. On one hand, the system \eqref{ndKP1}$\sim$\eqref{ndKP3} can be transformed to \eqref{bHM} under the transformation
\begin{equation}\label{Rel1}
U_{ij}=(a_i^{-1}-a_j^{-1})\frac{T_{ij}(\tau)\tau}{T_i(\tau)T_j(\tau)}.
\end{equation} 
On the other hand, the system \eqref{ndKP1}$\sim$\eqref{ndKP3} can be transformed to \eqref{fdkp} under the transformation
\begin{align}\label{Rel2}
U_{ij}=a_i^{-1}-a_{j}^{-1}+T_i(G)-T_j(G).
\end{align}

To make it clear, we will use the following picture to clarify the relations between the set of the bilinear equations and that of the nonlinear equations. 
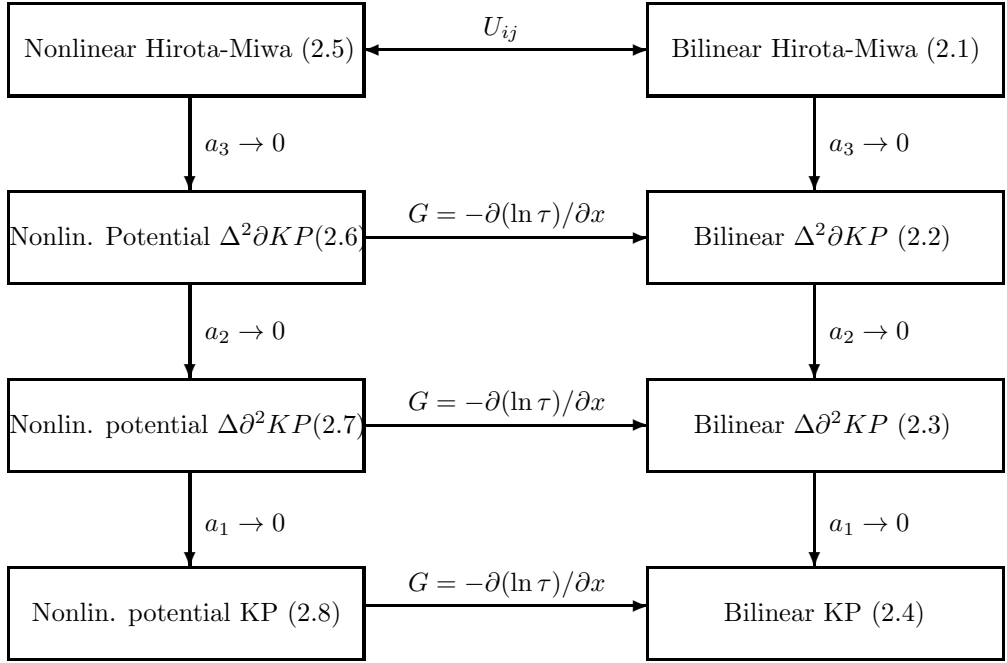
\begin{figure}[htb]
    \begin{center}
    \setlength{\unitlength}{1mm}
    \begin{picture}(90,85)
    \thicklines
      \put(-22,72){\framebox(47,12)}\put(25,78){\vector(1,0){38}}\put(63,78){\vector(-1,0){38}}\put(63,72){\framebox(47,12)}
      \put(-20,77){Nonlinear Hirota-Miwa \eqref{fdkp}}\put(41,80){$U_{ij}$}\put(66,77){Bilinear Hirota-Miwa \eqref{bHM}}
  \put(2,72){\vector(0,-1){13}} \put(85,72){\vector(0,-1){13}}
  \put(4,64.5){$a_3\rightarrow 0$}\put(87,64.5){$a_3\rightarrow 0$}
   \put(-22,47){\framebox(47,12)}\put(25,53){\vector(1,0){38}}\put(63,47){\framebox(47,12)}
         \put(-22,52){Nonlin. Potential $\Delta^2\partial KP$\eqref{nDPKP}}\put(31,55){$G=-\partial( \ln\tau)/ \partial x$}\put(69,52){Bilinear $\Delta^2\partial KP$ \eqref{bDPKP}}
   \put(2,47){\vector(0,-1){13}} \put(85,47){\vector(0,-1){13}} 
     \put(4,39.5){$a_2\rightarrow 0$}\put(87,39.5){$a_2\rightarrow 0$} 
       \put(-22,22){\framebox(47,12)}\put(25,28){\vector(1,0){38}}\put(63,22){\framebox(47,12)}
             \put(-22,27){Nonlin. potential $\Delta\partial^2 KP$\eqref{nPDKP}}\put(31,30){$G=-\partial( \ln\tau)/ \partial x$}\put(69,27){Bilinear $\Delta\partial^2 KP$ \eqref{bPDKP}}
    \put(2,22){\vector(0,-1){13}} \put(85,22){\vector(0,-1){13}} 
         \put(4,14){$a_1\rightarrow 0$}\put(87,14){$a_1\rightarrow 0$}
           \put(-22,-3){\framebox(47,12)}\put(25,4){\vector(1,0){38}}\put(63,-3){\framebox(47,12)}
     \put(-19,2){Nonlin. potential KP \eqref{nKP}}\put(31,6){$G=-\partial( \ln\tau)/ \partial x$}\put(73,2){Bilinear KP \eqref{bKP}}

%   \put(-9,86){\framebox(35,12)}\put(26,92){\vector(1,0){38}}\put(64,86){\framebox(35,12)}

%   \put(8,86){\vector(0,-1){13}} \put(82,86){\vector(0,-1){13}}
%   \put(-9,61){\framebox(35,12)}\put(26,67){\vector(1,0){38}}\put(64,61){\framebox(35,12)}
%    \put(8,61){\vector(0,-1){13}} \put(82,61){\vector(0,-1){13}} 
%       \put(-9,36){\framebox(35,12)}\put(26,42){\vector(1,0){38}}\put(64,36){\framebox(35,12)}
%    \put(8,36){\vector(0,-1){13}} \put(82,36){\vector(0,-1){13}} 
%           \put(-9,11){\framebox(35,12)}\put(26,17){\vector(1,0){38}}\put(64,11){\framebox(35,12)}
%    \put(8,36){\vector(0,-1){13}} \put(82,36){\vector(0,-1){13}} 
%   \put(-1,10){Nonlinear}
%     \put(-1,5){Hirota-Miwa}
% \put(-28,0){\framebox(23,10)}\put(-5,5){\vector(1,0){18}}
% \put(13,0){\framebox(23,10)}\put(36,5){\vector(1,0){18}}\put(54,0){\framebox(23,10)}\put(77,5){\vector(1,0){18}}
% \put(95,0){\framebox(23,10)}
% \put(-26,4){Hirota-Miwa}\put(19,4){$\Delta^2\partial$KP}\put(60,4){$\Delta\partial^2$KP}\put(104,4){KP}
%  \put(-2,6){$a_3\rightarrow 0$}\put(39,6){$a_2\rightarrow 0$}\put(80,6){$a_1\rightarrow 0$}
    \end{picture}
    \end{center}
    \caption{Continuum Limits and Connections\label{fig1}}
\end{figure}

\begin{rem}
By applying $\Delta_1$ to \eqref{nPDKP} and choosing $a_1=1$, $u=-\Delta_1 G$ and $y\rightarrow -t$, we have nothing but the differential-difference KP equation considered in \cite{DJM,TKT}
$$\Delta(u_t+2u_x-2uu_x)=(2+\Delta)u_{xx}.$$
\end{rem}

\begin{rem}
Under scaling transformations $u=-2G$ and $y\rightarrow -y,\, t\rightarrow -4t$, the nonlinear KP equation \eqref{nKP} can be transformed to the potential KP equation considered in \cite{GN3}
$$ (u_t+3u_x^2+u_{xxx})_x+3u_{yy}=0.$$
\end{rem}

%where $\Delta f_n=f_{n+1}-f_n$. 
%By the dependent variable transformation
%\begin{equation}
%u=\left(\ln (\tau_{n+1}/\tau_n)\right)_y,
%\end{equation}
%\begin{equation}
%(D_t+2D_y-D_y^2)\exp(D_n/2) \tau_n\cdot\tau_n=0
%\end{equation}

\section{Determinant solutions to the $\Delta^2\partial$KP equation}
In what follows, we will be mainly concerned with the $\Delta^2\partial$KP equation \eqref{bDPKP}, its determinant solutions proved by Wronskian (Casorati) technique and Pfaffians, and its corresponding coupled integrable system generated by Pfaffianization. 

\subsection{Casorati determinant solutions}
By the standard perturbation method, the semi-discrete KP equation \eqref{bDPKP} has the $2$-soliton solution
\begin{eqnarray}\label{tss}
&&\tau_2=1+\left(\frac{1-a_1q_1}{1-a_1p_1}\right)^{n_1}\left(\frac{1-a_2q_1}{1-a_2p_1}\right)^{n_2}\exp^{(p_1-q_1)x+const.}+\left(\frac{1-a_1q_2}{1-a_1p_2}\right)^{n_1}\left(\frac{1-a_2q_2}{1-a_2p_2}\right)^{n_2}\exp^{(p_2-q_2)x+const.}\nonumber\\
&&+\frac{(p_1-p_2)(q_1-q_2)}{(q_1-p_2)(p_1-q_2)}\left(\frac{(1-a_1q_1)(1-a_1q_2)}{(1-a_1p_1)(1-a_1p_2)}\right)^{n_1}\left(\frac{(1-a_2q_1)(1-a_2q_2)}{(1-a_2p_1)(1-a_2p_2)}\right)^{n_2}\exp^{(p_1-q_1+p_2-q_2)x+const.}.
\end{eqnarray}
Let $f_i(n_1,n_2,x)=e^{\xi_i}+e^{\hat{\xi}_i}$ with
\begin{eqnarray*}
&&e^{\xi_i}=(1-a_1p_i)^{-n_1}(1-a_2p_i)^{-n_2}\exp^{p_i x+const.},\\
&&e^{\hat{\xi}_i}=(1-a_1q_i)^{-n_1}(1-a_2q_i)^{-n_2}\exp^{q_i x+const.}.
\end{eqnarray*}
Consider the Wronkian of $f_1$ and $f_2$ with regard to the variable $x$, we have
\begin{align*}
\mbox{Wr}(f_1,f_2)&\equiv\begin{vmatrix}f_1&f_{1,x}\\f_2&f_{2,x}\end{vmatrix}=(q_2-q_1)\exp(\hat{\xi}_1+\hat{\xi}_2)\tau_2\\
&\propto 1+\exp{(\xi_1-\hat{\xi}_1)}+\exp{(\xi_2-\hat{\xi}_2)}+\frac{(p_1-p_2)(q_1-q_2)}{(p_1-q_2)(q_1-p_2)}\exp(\xi_1+\xi_2-\hat{\xi}_1-\hat{\xi}_2).
\end{align*}
It is not difficult to show that $\mbox{Wr}(f_1,f_2)$ also gives the 2-soliton solution to the $\Delta^2\partial$KP equation \eqref{bDPKP}. 

Actually, inspired by the $2$-soliton case above, we may expect that the semi-discrete KP equation \eqref{bDPKP} has the Wronskian determinant solution
\begin{align}\label{ws1}
\tau(n_1,n_2,x)=\mbox{Wr}(f_1,f_2,\cdots,f_N)=\begin{vmatrix}
f_1^{(0)}(n_1,n_2,x)&f_1^{(1)}(n_1,n_2,x)&\cdots&f_1^{(N-1)}(n_1,n_2,x)\\
f_2^{(0)}(n_1,n_2,x)&f_2^{(1)}(n_1,n_2,x)&\cdots&f_2^{(N-1)}(n_1,n_2,x)\\
\vdots&\vdots&\ddots&\vdots\\
f_N^{(0)}(n_1,n_2,x)&f_N^{(1)}(n_1,n_2,x)&\cdots&f_N^{(N-1)}(n_1,n_2,x)
\end{vmatrix}
\end{align}
where $f_i^{(m)}$ denotes $f_i^{(m)}={\partial^m f_i\over\partial x^m}$ and each $f_i\,\,(i=1,2,\dots,N-1)$ satisfies the dispersion relations
\begin{align}\label{dr1}
\nabla_m f_i={f_i(n_m)-f_i(n_m-1)\over a_m}={\partial f_i\over \partial x},\,\,\, \mbox{for}\, \,\,m=1,2.
\end{align}

Following Nimmo and Freeman's notation \cite{FN1,FN2}, we introduce some more compact notation and denote $\tau(n_1,n_2,x)$  as
\begin{align*}
\tau(n_1,n_2,x)=\begin{vmatrix}0_{\begin{smallmatrix}n_1\\n_2\end{smallmatrix}}&1_{\begin{smallmatrix}n_1\\n_2\end{smallmatrix}}&2_{\begin{smallmatrix}n_1\\n_2\end{smallmatrix}}&\dots&N-1_{\begin{smallmatrix}n_1\\n_2\end{smallmatrix}}\end{vmatrix}=\begin{vmatrix}
f_1^{(0)}(n_1,n_2,x)&f_1^{(1)}(n_1,n_2,x)&\cdots&f_1^{(N-1)}(n_1,n_2,x)\\
f_2^{(0)}(n_1,n_2,x)&f_2^{(1)}(n_1,n_2,x)&\cdots&f_2^{(N-1)}(n_1,n_2,x)\\
\vdots&\vdots&\ddots&\vdots\\
f_N^{(0)}(n_1,n_2,x)&f_N^{(1)}(n_1,n_2,x)&\cdots&f_N^{(N-1)}(n_1,n_2,x)
\end{vmatrix}.
\end{align*}
We can make a further notational simplification by suppressing the indices where the variables are unshifted to leave
$$\tau(n_1,n_2,x)=\begin{vmatrix}0&1&2&\dots&N-1\end{vmatrix}.$$
By resorting to the dispersion relations \eqref{dr1}, for $i=1,2$, we have
\begin{align*}
&\tau(n_i+1)=\begin{vmatrix}0&1&\dots&N-2&N-1_{n_i+1}\end{vmatrix},\\
&a_i\tau(n_i+1)=\begin{vmatrix}0&1&\dots&N-2&N-2_{n_i+1}\end{vmatrix},\\
&(a_1-a_2)\tau(n_1+1,n_2+1)=\begin{vmatrix}0&1&\dots&N-3&N-2_{n_2+1}&N-2_{n_1+1}\end{vmatrix},\\
&\tau_x(n_i+1)=\begin{vmatrix}0&1&\dots&N-3&N-1&N-1_{n_i+1}\end{vmatrix}+\begin{vmatrix}0&1&\dots&N-3&N-2&N_{n_i+1}\end{vmatrix},\\
&a_i{\tau_x(n_i+1)}=\begin{vmatrix}0&1&\dots&N-3&N-1&N-2_{n_i+1}\end{vmatrix}+\begin{vmatrix}0&1&\dots&N-3&N-2&N-1_{n_i+1}\end{vmatrix}.
\end{align*}
By using the above-mentioned expressions, the semi-discrete KP equation \eqref{bDPKP} reduces to the Pl$\ddot{u}$cker relation
\begin{align*}
&(a_1\tau_x(n_1+1))(a_2\tau(n_2+1))-(a_2\tau_x(n_2+1))(a_1\tau(n_1+1))\\
&\quad+(a_1\tau(n_1+1))\tau(n_2+1)-\tau(n_1+1)(a_2\tau(n_2+1))-\tau((a_1-a_2)\tau(n_1+1,n_2+1))\\
&=\begin{vmatrix}0&1&\dots&N-2&N-2_{n_2+1}\end{vmatrix}\begin{vmatrix}0&1&\dots&N-3&N-1&N-2_{n_1+1}\end{vmatrix}\\
&\quad -\begin{vmatrix}0&1&\dots&N-2&N-1\end{vmatrix}\begin{vmatrix}0&1&\dots&N-3&N-2_{n_2+1}&N-2_{n_1+1}\end{vmatrix}\\
&\quad-\begin{vmatrix}0&1&\dots&N-2&N-2_{n_1+1}\end{vmatrix}\begin{vmatrix}0&1&\dots&N-3&N-1&N-2_{n_2+1}\end{vmatrix}\\
&\equiv0.
\end{align*}
Therefore, we have proved that the Wronskian determinant $\tau(n_1,n_2,x)$ given by \eqref{ws1} satisfies \eqref{bDPKP}.

\subsection{Grammian determinant solutions}
By rewriting the $2$-soliton solution $\tau_2$ given by \eqref{tss}, we have
\begin{align*}
&\tau_2=1+{1\over p_1-q_1}\exp{(\xi_1-\hat{\xi}_1)}+{1\over p_2-q_2}\exp{(\xi_2-\hat{\xi}_2)}\\
&\qquad+{(p_1-p_2)(q_1-q_2)\over(p_1-q_2)(q_1-p_2) (p_1-q_1)(p_2-q_2)}\exp{(\xi_1-\hat{\xi}_1+\xi_2-\hat{\xi}_2)}\nonumber\\
&\quad=\begin{vmatrix}
1+{1\over p_1-q_1}\exp{(\xi_1-\hat{\xi}_1)}&{1\over p_1-q_2}\exp{(\xi_1-\hat{\xi}_2)}\\
{1\over p_2-q_1}\exp{(\xi_2-\hat{\xi}_1)}&1+{1\over p_2-q_2}\exp{(\xi_2-\hat{\xi}_2)}
\end{vmatrix}
\end{align*}
which can be reinterpreted as a Grammian determinant
\begin{align*}
\tau_2=\mbox{det}(a_{ij})_{1\le i,j\le 2},\,\, a_{ij}=\delta_{ij}+\int^x \exp(\xi_i-\hat{\xi}_j)dx.
\end{align*}

We may then expect that the semi-discrete KP equation \eqref{bDPKP} has the $N\times N$ Grammian determinant solution 
\begin{align}
\tau_{n_1}^{(n_2)}=\mbox{det}(a_{ij})_{1\le i,j\le N},\,\, a_{ij}=c_{ij}+\int^x f_i(n_1,-n_2,x) g_j(-n_1,n_2,x) dx, \,\, c_{ij}=const.\label{gs}
\end{align}
with $f_i(n_1,-n_2,x)$ and $g_j(-n_1,n_2,x)$ satisfying the dispersion relations
\begin{align}
&\nabla_1 f_i={f_i(n_1)-f_i(n_1-1)\over a_1}={\partial f_i\over \partial x}, \qquad \nabla_2 f_i={f_i(-n_2)-f_i(-n_2+1)\over a_2}={\partial f_i\over\partial x},\label{dr21}\\
&\Delta_2 g_j ={g_j(n_2+1)-g_j(n_2)\over a_2}={\partial g_j\over \partial x},\qquad\Delta_2 g_j= {g_j(-n_1-1)-g_j(-n_1)\over a_1}={\partial g_j\over \partial x}.\label{dr22}
\end{align}

In order to prove that \eqref{gs} does give the Grammian determinant solution to the semi-discrete KP equation \eqref{bDPKP}, we need to calculate the derivatives and the shifts of $\tau_{n_1}^{(n_2)}$. Let us express $\tau_{n_1}^{(n_2)}$ by means of a pfaffian as 
\begin{align}
\tau_{n_1}^{(n_2)}=(1,2,\dots,N,N^*,\dots,2^*,1^*)_{n_1}^{(n_2)}
\end{align}
where $(i,j^*)_{n_1}^{(n_2)}=c_{ij}+\int^x f_i(n_1,-n_2,x) g_j(-n_1,n_2,x) dx,\,\, c_{ij}=const.$ and $(i,j)=(i^*,j^*)=0$.

 Next, let us introduce pfaffian entries
 \begin{align}
 & (d_m,j^*)=g_j(-n_1,n_2+m),\, (d_m^*,i)=f_i(n_1,-n_2+m), \,(c_m^*,i)=f_i(n_1+m,-n_2),\\
 &(c_m^*,d_n^*)=0,\,(c_m^*,d_m)=0,\,(d_m,d_n^*)=0,\, (d_m,i)=0, \,(d_m^*,j^*)=0, \,(c_m^*,j^*)=0.
 \end{align} 
By using the dispersion relations \eqref{dr21}$\sim$\eqref{dr22}, we have
\begin{align*}
&(i,j^*)_{n_1+1}^{(n_2)}=(i,j^*)_{n_1}^{(n_2)}+a_1f_i(n_1+1)g_j=(i,j^*)_{n_1}^{(n_2)}+a_1(d_0,c_1^*,i,j^*)_{n_1}^{(n_2)},\\
&(i,j^*)_{n_1}^{(n_2+1)}=(i,j^*)_{n_1}^{(n_2)}+a_2 f_i(-n_2-1)g_j=(i,j^*)_{n_1}^{(n_2)}+a_2(d_0,d_{-1}^*,i,j^*)_{n_1}^{(n_2)},
\end{align*}
\begin{align*}
& (i,j^*)_{n_1,x}^{(n_2+1)}=f_i(-n_2-1)g_j(n_2+1)=(d_1,d_{-1}^*,i,j^*),\\
&(i,j^*)_{n_1+1,x}^{(n_2)}=f_i(n_1+1)g_j(-n_1-1)=a_2^{-1}f_i(n_1+1)(a_1g_j(n_2+1)-(a_1-a_2)g_j)\\
&\qquad\qquad\quad=a_2^{-1}(a_1(d_1,c_1^*,i,j^*)-(a_1-a_2)(d_0,c_1^*,i,j^*)),\\
&(i,j^*)_{n_1+1}^{(n_2+1)}=(i,j^*)_{n_1}^{(n_2)}+a_2f_i(-n_2-1)g_j+a_1(a_1-a_2)^{-1}g_j(n_2+1)(a_1f_i(n_1+1)-a_2f_i(-n_2-1))\\
&\qquad\qquad\quad=(i,j^*)_{n_1}^{(n_2)}+a_2(d_0,d_{-1}^*,i,j^*)+a_1(a_1-a_2)^{-1}(a_1(d_1,c_1^*,i,j^*)-a_2(d_1,d_{-1}^*,i,j^*)).
\end{align*}
By using the addition formulae and derivative formulae for pfaffians, we have
\begin{align*}
&\tau_{n_1+1}^{(n_2)}=\tau_{n_1}^{(n_2)}+a_1(d_0,c_1^*, \bullet),\,\,\tau_{n_1}^{(n_2+1)}=\tau_{n_1}^{(n_2)}+a_2(d_0,d_{-1}^*,\bullet),\\
&\tau_{n_1,x}^{(n_2+1)}=(d_1,d_{-1}^*,\bullet),\,\,\tau_{n_1+1,x}^{(n_2)}=a_1a_2^{-1}(d_1,c_1^*,\bullet)+(1-a_1a_2^{-1})(d_0,c_1^*,\bullet),\\
&\tau_{n_1+1}^{(n_2+1)}=\tau_{n_1}^{(n_2)}+a_2(d_0,d_{-1}^*,\bullet)+a_1(a_1-a_2)^{-1}(a_1(d_1,c_1^*,\bullet)-a_2(d_1,d_{-1}^*,\bullet)+a_1a_2(d_1,c_1^*,d_0,d_{-1}^*,\bullet)),
\end{align*}
where we have used $\bullet$ to represent the list of indices $1,2,\dots,N,N^*,\dots,2^*,1^*$ common to each pfaffian.

By substituting the above expressions, the semi-discrete KP equation \eqref{bDPKP} reduces to nothing but the Jacobi identity for determinants
\begin{align*}
&(a_1-a_2)(T_1(\tau)T_2(\tau)-T_{12}(\tau)\tau)+a_1a_2D_xT_1(\tau)\cdot T_2(\tau)\\
&=-a_1^2a_2[((d_1,c_1^*,d_0,d_{-1}^*,\bullet))(\bullet)-(d_0,d_{-1}^*,\bullet)(d_1,c_1^*,\bullet)+(d_1,d_{-1}^*,\bullet)(d_0,c_1^*,\bullet)]\\
&\equiv0.
\end{align*}
We have proved the Grammian determinant solutions \eqref{gs} to the semi-discrete KP equation \eqref{bDPKP}.

\begin{rem}
One can refer to \cite{HR1,HR2,HRN} for definitions of Pfaffians, their properties and their applications in soliton theory and integrable systems.
\end{rem}

\section{The coupled integrable system for the $\Delta^2\partial$KP equation}
In this section, we shall use the Pfaffianization procedure to seek the coupled integrable system of the semi-discrete KP equation \eqref{bDPKP}. For this purpose, we require Pfaffians with elements satisfying the Pfaffianized form of the dispersion relations \eqref{dr1}. Therefore, the entries in our Pfaffians are chosen to satisfy
\begin{align}
&\nabla_{n_1}(i,j)_{n_1}^{(n_2)}=a_1^{-1}((i,j)_{n_1}^{(n_2)}-(i,j)_{n_1-1}^{(n_2)})=(i,j+1)_{n_1}^{(n_2)}+(i+1,j)_{n_1}^{(n_2)}-a_1(i+1,j+1)_{n_1}^{(n_2)},\label{pdr1}\\
&\nabla_{n_2}(i,j)_{n_1}^{(n_2)}=a_2^{-1}((i,j)_{n_1}^{(n_2)}-(i,j)_{n_1}^{(n_2-1)})=(i,j+1)_{n_1}^{(n_2)}+(i+1,j)_{n_1}^{(n_2)}-a_2(i+1,j+1)_{n_1}^{(n_2)},\label{pdr2}\\
&(i,j)_{n_1,x}^{(n_2)}=(i+1,j)_{n_1}^{(n_2)}+(i,j+1)_{n_1}^{(n_2)}.\label{pdr3}
\end{align}
or equivalently
\begin{align}
&(i,j)_{n_1-1}^{(n_2)}=(i,j)_{n_1}^{(n_2)}-a_1(i,j+1)_{n_1}^{(n_2)}-a_1(i+1,j)_{n_1}^{(n_2)}+a_1^2(i+1,j+1)_{n_1}^{(n_2)},\\
&(i,j)_{n_1}^{(n_2-1)}=(i,j)_{n_1}^{(n_2)}-a_2(i,j+1)_{n_1}^{(n_2)}-a_2(i+1,j)_{n_1}^{(n_2)}+a_2^2(i+1,j+1)_{n_1}^{(n_2)},\\
&(i,j)_{n_1,x}^{(n_2)}=(i+1,j)_{n_1}^{(n_2)}+(i,j+1)_{n_1}^{(n_2)}.
\end{align}

If we wish to consider solutions of this coupled system then we need entries in Pfaffians to satisfy the dispersion relations \eqref{pdr1}$\sim$\eqref{pdr3} introduced earlier. By refering to \cite{GNT}, we can choose
\begin{align*}
(i,j)_{n_1}^{(n_2)}=\sum_{m=1}^M[f_{2m-1}(i)f_{2m}(j)-f_{2m-1}(j)f_{2m}(i)]
\end{align*}
where $f_m$ satisfy the equations
\begin{align*}
&f_m(n_1,n_2,x: i)-f_m(n_1-1,n_2,x: i)=a_1f_m(n_1,n_2,x: i+1),\\
&f_m(n_1,n_2,x: i)-f_m(n_1,n_2-1,x: i)=a_2 f_m(n_1,n_2,x: i+1),\\
&{\partial\over \partial x}f_m(n_1,n_2,x: i)=f_m(n_1,n_2,x: i+1).
\end{align*}

Let us take 
\begin{align}
\tau_{n_1}^{(n_2)}=(1,2,\dots,N)_{n_1}^{(n_2)},\,\, N\,\,\,  \mbox{even}\label{wg1}
\end{align}
together with $(i,c_j)_{n_1}^{(n_2)}=a_j^{N+1-i}$ and $(c_i,c_j)=0$ for $i\neq j$, then we have
\begin{align}
&\tau_{n_1-1}^{(n_2)}=(1,\dots,N,N+1,c_1)_{n_1}^{(n_2)},\label{wg2}\\
&\tau_{n_1}^{(n_2-1)}=(1,\dots,N,N+1,c_2)_{n_1}^{(n_2)},\\
&\tau_{n_1-1}^{(n_2-1)}={a_1a_2\over a_2-a_1}(1,\dots,N,N+1,N+2,c_1,c_2)_{n_1}^{(n_2)},\\
&{\partial\over\partial x}\tau_{n_1-1}^{(n_2)}=(1,\dots,N,N+2,c_1)_{n_1}^{(n_2)}-a_1^{-1}(1,\dots,N,N+1,c_1)_{n_1}^{(n_2)},\\
&{\partial\over\partial x}\tau_{n_1}^{(n_2-1)}=(1,\dots,N,N+2,c_2)_{n_1}^{(n_2)}-a_2^{-1}(1,\dots,N,N+1,c_2)_{n_1}^{(n_2)},\\
&{\partial\over\partial x}\tau_{n_1-1}^{(n_2-1)}={a_1a_2\over (a_2-a_1)}(1,\dots,N+1,N+3,c_1,c_2)_{n_1}^{(n_2)}-{a_1+a_2\over a_2-a_1}(1,\dots,N+2,c_1,c_2)_{n_1}^{(n_2)}.\label{wg3}
\end{align}

Before we move to the next step,  let us first review two simple Pfaffian identites:
\begin{align}
&(1,\dots,N,\alpha,\beta,\gamma,\delta)(1,\dots,N)\nonumber\\
&\qquad-(1,\dots,N,\alpha,\beta)(1,\dots,N,\gamma,\delta)\nonumber\\
&\qquad+(1,\dots,N,\alpha,\gamma)(1,\dots,N,\beta,\delta)\nonumber\\
&\qquad-(1,\dots,N,\alpha,\delta)(1,\dots,N,\beta,\gamma)=0,\,\, N\,\,\, \mbox{even}\\
&(1,\dots,N,\alpha,\beta,\gamma)(1,\dots,N,\delta)\nonumber\\
&\qquad-(1,\dots,N,\alpha,\beta,\delta)(1,\dots,N,\gamma)\nonumber\\
&\qquad+(1,\dots,N,\alpha,\gamma,\delta)(1,\dots,N,\beta)\nonumber\\
&\qquad-(1,\dots,N,\delta,\gamma,\delta)(1,\dots,N,\alpha)=0,\,\, N\,\,\, \mbox{ood}.
\end{align}
Actually, these two Pfaffian identities are generalizations of the Pl$\ddot{u}$cker relation and Jacobi identities for determinants, respectively. This is the reason that we can extend the semi-discrete KP equation \eqref{bDPKP} to its coupled system.

Following Hirota and Ohta's procedure \cite{HO,HR1}, we now introduce two new functions 
\begin{align}
&\sigma_{n_1}^{(n_2)}=(1,\dots,N,N+1,N+2)_{n_1}^{(n_2)},\label{wg4}\\
&\tilde{\sigma}_{n_1}^{(n_2)}=(1,\dots,N-2)_{n_1}^{(n_2)}.\label{wg5}
\end{align}
We can show that $\tau_{n_1}^{(n_2)}, \sigma_{n_1}^{(n_2)}$ and $\tilde{\sigma}_{n_1}^{(n_2)}$ satisfy the following bilinear equations
\begin{align}
&a_1a_2(\tau_{n_1-1}^{(n_2)}\tau_{n_1,x}^{(n_2-1)}-\tau_{n_1-1,x}^{(n_2)}\tau_{n_1}^{(n_2-1)})\nonumber\\
&+(a_1-a_2)(\tau_{n_1-1}^{(n_2)}\tau_{n_1}^{(n_2-1)}-\tau\tau_{n_1-1}^{(n_2-1)})-(a_2-a_1)\sigma_{n_1}^{(n_2)}\tilde{\sigma}_{n_1-1}^{(n_2-1)}=0,\label{cs1}\\
&a_1a_2(a_2-a_1)[\tau_{n_1}^{(n_2)}\tilde{\sigma}_{n_1-1,x}^{(n_2-1)}-\tau_{n_1,x}^{(n_2)}\tilde{\sigma}_{n_1-1}^{(n_2-1)}]\nonumber\\
&+a_1^2a_2^2(\tilde{\sigma}_{n_1-1}^{(n_2)}\tau_{n_1}^{(n_2-1)}-\tilde{\sigma}_{n_1}^{(n_2-1)}\tau_{n_1-1}^{(n_2)})+(a_2^2-a_1^2)\tau_{n_1}^{(n_2)}\tilde{\sigma}_{n_1-1}^{(n_2-1)}=0,\label{cs2}\\
&a_1a_2(a_2-a_1)[\sigma_{n_1}^{(n_2)}\tau_{n_1-1,x}^{(n_2-1)}-\sigma_{n_1,x}^{(n_2)}\tau_{n_1-1}^{(n_2-1)}]\nonumber\\
&+a_1^2a_2^2(\tau_{n_1-1}^{(n_2)}\sigma_{n_1}^{(n_2-1)}-\tau_{n_1}^{(n_2-1)}\sigma_{n_1-1}^{(n_2)})+(a_2^2-a_1^2)\sigma_{n_1}^{(n_2)}\tau_{n_1-1}^{(n_2-1)}=0.\label{cs3}
\end{align}
%\begin{align*}
%&(a_1-a_2)(\tau_{n_1-1}^{(n_2)}\tau_{n_1}^{(n_2-1)}-\tau\tau_{n_1-1}^{(n_2-1)})+a_1a_2(\tau_{n_1-1}^{(n_2)}\tau_{n_1,x}^{(n_2-1)}-\tau_{n_1-1,x}^{(n_2)}\tau_{n_1}^{(n_2-1)})\\
%&=a_1a_2(1,\dots,N)_{n_1}^{(n_2)}(1,\dots,N,N+1,N+2,c_1,c_2)_{n_1}^{(n_2)}\\
%&\quad-(1,\dots,N,N+1,c_2)_{n_1}^{(n_2)}(1,\dots,N,N+2,c_1)_{n_1}^{(n_2)}\\
%&\quad+(1,\dots,N,N+1,c_1)_{n_1}^{(n_2)}(1,\dots,N,N+2,c_2)]_{n_1}^{(n_2)}\\
%&=a_1a_2(1,\dots,N,N+1,N+2)_{n_1}^{(n_2)}(1,\dots,N,c_1,c_2)_{n_1}^{(n_2)}\\
%&=(a_2-a_1)(1,\dots,N,N+1,N+2) {a_1a_2\over a_2-a_1}(1,\dots,N,c_1,c_2)\\
%&=(a_2-a_1)(1,\dots,N,N+1,N+2)_{n_1}^{(n_2)}(1,2,\dots,N-2)_{n_1-1}^{(n_2-1)}.
%\end{align*}
In fact, substitution of \eqref{wg1}$\sim$\eqref{wg3} and \eqref{wg4}$\sim$\eqref{wg5} into \eqref{cs1} leads to the following Pfaffian identity
\begin{align*}
&a_1a_2(1,\dots,N,N+1,N+2,c_1,c_2)(1,\dots,N)\nonumber\\
&\qquad-a_1a_2(1,\dots,N,N+1,N+2)(1,\dots,N,c_1,c_2)\nonumber\\
&\qquad+a_1a_2(1,\dots,N,N+1,c_1)(1,\dots,N,N+2,c_2)\nonumber\\
&\qquad-a_1a_2(1,\dots,N,N+1,c_2)(1,\dots,N,N+2,c_1)=0.
\end{align*}

In the same way, one can easily show that \eqref{cs2} and \eqref{cs3} are equivalent to the following two Pfaffian identities, respectively.
\begin{align}
&a_1^2a_2^2(1,\dots,N-1,N)(1,\dots,N-1,N+1,c_1,c_2)\nonumber\\
&\qquad-a_1^2a_2^2(1,\dots,N-1,N+1)(1,\dots,N-1,N,c_1,c_2)\nonumber\\
&\qquad+a_1^2a_2^2(1,\dots,N-1,c_1)(1,\dots,N-1,N,N+1,c_2)\nonumber\\
&\qquad-a_1^2a_2^2(1,\dots,N-1,c_2)(1,\dots,N-1,N,N+1,c_1)=0,\\
&a_1^2a_2^2(1,\dots,N+1,N+2)(1,\dots,N+1,N+3,c_1,c_2)\nonumber\\
&\qquad-a_1^2a_2^2(1,\dots,N+1,N+3)(1,\dots,N+1,N+2,c_1,c_2)\nonumber\\
&\qquad+a_1^2a_2^2(1,\dots,N+1,c_1)(1,\dots,N+1,N+2,N+3,c_2)\nonumber\\
&\qquad-a_1^2a_2^2(1,\dots,N+1,c_2)(1,\dots,N+1,N+2,N+3,c_1)=0.
\end{align}

\begin{rem}
We have successfully found the coupled integrable system \eqref{cs1}$\sim$\eqref{cs3} for the semi-discrete KP equation \eqref{bDPKP} through Pfaffianization. The essential step is to generalize the Wronkian determinant solutions for the semi-discrete KP equation \eqref{bDPKP} to the Wronskian-type Pfaffians which no longer satisfy \eqref{bDPKP} but its coupled systems  \eqref{cs1}$\sim$\eqref{cs3}. Since the semi-discrete KP equation \eqref{bDPKP} also has Grammian determinant solutions, similarly, we can generalize the Grammian determinant solutions to the Grammian-type Pfaffians to derive the same coupled system \eqref{cs1}$\sim$\eqref{cs3}. In other words, the coupled system \eqref{cs1}$\sim$\eqref{cs3} has Grammian-type Pfaffian solutions too which we will not give here. 
\end{rem}

\section{Conclusion}
The well-known Hirota-Miwa equation is very fundamental in soliton theory and integrable systems. Many integrable systems can be generated from it. In \cite{HR,MT}, by taking continuum limits of the Hirota-Miwa equation under the Miwa transformation, the semi-discrete KP equation, the differential-difference KP equation and the well-known KP equation were derived. In our previous work \cite{LNT}, we found a remarkable nonlinear fully discrete KP equation expressed in terms of a single equation. This nonlinear Hirota-Miwa equation can be also taken continuum limits to derive the nonlinear semi-discrete KP equation, the nonlinear differential-difference equation and the nonlinear KP equation under the Miwa transformation. In this paper, we establish connections between these two cascades of integrable systems. Later on, we are mainly concerned with the semi-screte bilinear KP equation which is seldomly studied. We present both of its Casorati determinant solutions and Grammian determinant solutions. By using the Pfaffianization procedure proposed by Hirota and Ohta, we are able to derive the couple integrable system for the semi-discrete KP equation.

\section*{Acknowledgement}
This work was supported by the National Natural Science Foundation of China (Grant No. 11271266 and 11371323) and the China Scholarship Council. One of the authors (C.X. Li) would like to thank Professor Xing-Biao Hu and Professor Alex Kasman for valuable discussions. Dr. Li would also like to thank for the hospitality of the Department of Mathematics during her visit to the College of Charleston.

\end{document}